# Spatiotemporal Adaptive Quantization for Video Compression Applications

Lee Prangnell

Department of Computer Science, University of Warwick, England, UK

**ABSTRACT** — JCT-VC HEVC HM 16 includes a Coding Unit (CU) level adaptive Quantization Parameter (QP) technique named AdaptiveQP. It is designed to perceptually adjust the QP in Y, Cb and Cr Coding Blocks (CBs) based only on the variance of samples in a luma CB. In this paper, we propose an adaptive quantisation technique that consists of two contributions. The first contribution relates to accounting for the variance of chroma samples, in addition to luma samples, in a CU. The second contribution relates to accounting for CU temporal information as well as CU spatial information. Moreover, we integrate into our method a lambda refined QP technique to reduce complexity associated multiple QP optimizations in the Rate Distortion Optimization process. We evaluate the proposed technique on 4:4:4, 4:2:2, 4:2:0 and 4:0:0 YCbCr test sequences, for which we quantify the results using the Bjøntegaard Delta Rate (BD-Rate) metric. Our method achieves a maximum BD-Rate reduction of 23.1% (Y), 26.7% (Cr) and 25.2% (Cb). Furthermore, a maximum encoding time reduction of 4.4% is achieved.

## 1. INTRODUCTION

In the JCT-VC HEVC Test Model 16 reference software (HM 16), a CU level QP technique (AdaptiveQP) has been adopted, which is based on a technique that is integrated into MPEG-2 Test Model 5 [1]. The AdaptiveQP tool is an adaptive quantization method that modifies the QP of a $2N \times 2N$ CU according to luma activity, which is quantified by the variance of the luma samples contained within $N \times N$ CU sub-blocks [1]. This technique is designed to exploit Human Visual System (HVS) spatial masking. Moreover, it has been shown to provide coding efficiency improvements in comparison with Uniform Reconstruction Quantization (URQ). With the Adaptive QP tool, the QP is decreased for areas where there is low spatial activity of luma samples. Conversely, the QP in increased for areas where there is high spatial activity of luma samples. This technique takes into account only the variance of luma samples; in addition, it does not account for motion information in a CU, thus leaving room for improvement.

Adaptive quantization methods similar to the AdaptiveQP tool in HM 16 have been previously proposed. The work in [2] proposes a technique which is designed to adaptively adjust the quantization step size by exploiting intensity masking of the HVS; this technique has its roots in the Just Noticeable Distortion (JND) model of lossy compression. Similar to the AdaptiveQP tool in HM 16, this technique concentrates on luma information; therefore, the authors of this technique evaluate the method on the luma component of sequences. Moreover, this technique does not take into account motion information; consequently, the temporal masking phenomenon of the HVS is not exploited. The proposed technique in [3] is a transform coefficient level technique that quantizes coefficients individually in a Transform Block (TB). In comparison with URQ, it reduces the quantization step size for low frequency transform coefficients. Similar to the technique proposed in [2], this method does not take into account motion.

In this paper, we propose a CU level adaptive quantization technique (ACUQ) to improve upon the AdaptiveQP tool. As summarized below, ACUQ includes two novel contributions and the integration of a lambda QP refinement technique.

*Chroma Cb and Cr Data*

In addition to accounting for the variance of the luma samples contained within a CU, ACUQ also fully accounts for the variance of the chroma samples. Due to the advent of contemporary consumer electronics visual display technologies that support ITU-R Recommendation BT.2020-2 RGB and YCbCr 4:4:4 color video data formats [4], we contend that it is necessary to account for CU chroma information in order to attain a more accurate reflection of CU spatial activity in 4:4:4, 4:2:2 and 4:2:0 video data. The Largest Coding Unit (LCU) in HEVC is $64 \times 64$ samples — CU QuadTree (QT) Depth Level = 0. The Smallest Coding Unit (SCU) is $8 \times 8$ samples — CU QuadTree (QT) Depth Level = 3. Therefore, each CU potentially supports up to $64 \times 64$ luma samples and $64 \times 64$ chroma samples assuming no chroma subsampling (i.e., YCbCr 4:4:4 video data).

*Temporal Masking*

ACUQ accounts for motion information in a picture in order to exploit the temporal masking phenomenon of the HVS. A multitude of psychophysical experiments confirm that temporal masking is a well established phenomenon of the HVS [5, 6]. In ACUQ, temporal masking is achieved by utilizing a threshold value quantified by the arithmetic mean motion vector magnitude within an entire frame. If the magnitude of a motion vector in a Prediction Unit (PU) exceeds this threshold, then the region is considered to be high motion. Subsequently, the QP value is incremented in this temporally intense region. The advantage of the temporal masking component of ACUQ is twofold. First, a higher QP in high motion regions potentially results in an overall decreased bitrate. Second, the increased distortion in high motion areas, incurred by the incremented QP value, is not perceptible. Therefore, this equates to bitrate savings without affecting the visual quality the reconstructed video data.

*Lamda QP Refinement*

In the RDO process, instead of utilizing the default multiple QP optimization method to improve R-D performance, refining the QP according to the Lagrange multiplier (lambda) has shown benefits in terms of significantly decreasing encoding times while also attaining small BD-Rate reductions of approximately 1.7% [7]. Multiple QP optimization refers to a computationally expensive RDO technique for improving coding efficiency with respect to selecting the most appropriate QP in the quantization process. We integrate the technique proposed in [7] into ACUQ for the purpose of decreasing encoding times.

The rest of this paper is organized as follows. Section 2 briefly reviews the AdaptiveQP tool. Section 3 includes detailed information on the proposed ACUQ method. Section 4 includes the evaluations, results and discussion. Finally, Section 5 concludes this paper.

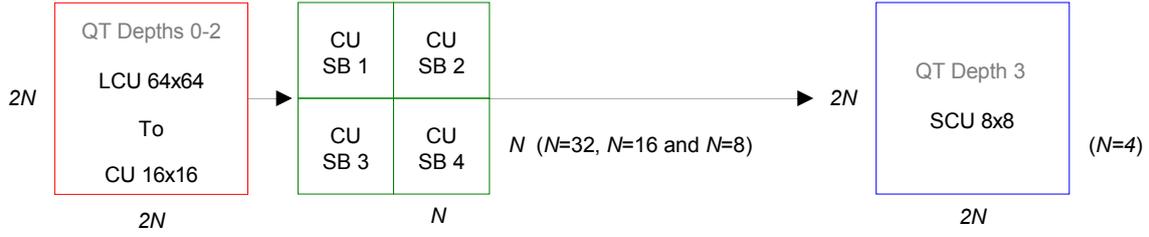

**Fig. 1.** The HEVC standard supports CUs of size 2N×2N and CU sub-blocks of size N×N [8, 9]. The Largest Coding Unit (LCU) at CU QuadTree (QT) Depth Level = 0 supports up to 64×64 luma and chroma samples (as shown in red). The CUs at QT depth levels 0-2 are partitioned into CU sub-blocks of size N×N supporting from 32×32 to 8×8 luma and chroma samples (as shown in green), respectively. There is no partitioning in the 2N×2N Smallest Coding Unit (SCU), as shown in blue. The AdaptiveQP tool operates at CU QT depth levels 0-2.

## 2. ADAPTIVE QP TOOL IN HEVC

The AdaptiveQP tool in HM 16 modifies the QP at the CU level based on the activity in a CU of size 2N×2N. The CU activity is determined by the variance of luma samples in the four constituent N×N sub-blocks. This technique operates at CU QT depth levels 0-2 (see Fig. 1). Similar to the temporal masking phenomenon of the HVS, in the spatial domain human beings are perceptually less sensitive to quantization-induced compression artifacts in regions where there exists significant luma sample variation. Therefore, the AdaptiveQP tool applies a lower QP value to regions in which there is low luma spatial activity. Conversely, a higher QP is utilized for regions in which there exists high luma spatial activity, thus achieving bitrate savings without affecting visual quality. The CU level adaptive QP, denoted by $Q$, in the AdaptiveQP tool is computed in (1) [1]:

$$Q = QP + \left[6 \times \log_2(R)\right] \quad (1)$$

where $QP$ corresponds to the slice level QP and where $R$ refers to the normalized spatial activity of the 2N×2N CU. $R$ is computed in (2):

$$R = \frac{(s \times (Y')) + t_p}{(Y') + (s \times t_p)} \quad (2)$$

where parameter $s$ is a scaling factor associated with the QP adaptation range $A$. Parameter $A$ corresponds to the maximum offset allowed for the QP value in the CU, where the default value is $A = 6$. $Y'$ corresponds to the luma spatial activity of a 2N×2N CU and parameter $t_p$ refers to the average activity for all 2N×2N CUs in picture $p$. Parameter $s$ is shown in (3) and $Y'$ is computed in (4):

$$s = 2^{\frac{A}{6}} \quad (3)$$

$$Y' = 1 + \min\left(Y^2(sb)\right) \quad \text{where } sb = 1,...,4 \quad (4)$$

where $Y^2(sb)$ denotes the luma sample variance for an N×N CU sub-block, denoted as $sb$. Consequently, $Y^2$ is quantified as the variance of luma pixel intensities, which is computed in (5):

$$Y^2 = \frac{1}{N^2} \sum_{j=1}^{N^2} (LS_j - a)^2 \quad (5)$$

where $N^2$ corresponds to the product of luma samples in an N×N luma CU sub-block; N=8, N=16 or N=32 depending on the QT depth level (see Fig. 1). $LS$ refers to the luma sample values in the $j^{th}$ CU N×N sub-block and $a$ denotes the mean sample intensity of the N×N luma CU sub-block, which is computed in (6).

$$a = \frac{1}{N^2} \sum_{j=1}^{N^2} LS_j \quad (6)$$

## 3. PROPOSED ACUQ TECHNIQUE

As previously described in Section 1, the proposed ACUQ method consists of three components, the first two of which represent our contributions. 1) Accounting for chroma information in a CU in addition to luma information. 2) Quantifying the motion information in a CU to exploit the temporal masking phenomenon of the HVS. 3) Integrating the lambda QP refinement technique in [7] to bypass multiple QP optimization and, thus, decreasing encoding times.

Let us denote $\tilde{Q}$ for computing the CU level QP as follows:

$$\tilde{Q} = \begin{cases} (D+q) + \left[(6\log_2(X))\right] & \text{if YCbCr} \neq 4:0:0; \\ (D+q) + \left[(6\log_2(R))\right] & \text{if YCbCr} = 4:0:0. \end{cases} \quad (7)$$

where $D$ corresponds to a parameter that increments $q$, $X$ denotes the normalized spatial activity of a 2N×2N CU, which accounts for both luma and chroma information and $q$ corresponds to the QP value derived from the technique in [7]. $D$ is computed in (8):

$$D = \begin{cases} 1 & \text{if } M > MVM; \\ 0 & \text{otherwise.} \end{cases} \quad (8)$$

where $M$ corresponds to the magnitude of motion vector $MV$ in a PU and where $MVM$ is a function that adaptively computes the arithmetic mean motion vector magnitude in a PU of an entire frame. $MVM$ is, thus, in place for the purpose of quantifying an adaptive threshold value which the magnitude $M$ of motion vector $MV$ must exceed in order for a region to be considered as high motion. $D$ is employed to increment $q$, thus potentially resulting in bitrate reductions without incurring a perceptible loss in visual quality. $MVM$ is computed in (9) and $M$ is computed in (10):

$$MVM(n) = \frac{1}{C} \sum_{i=1}^{C} M(n,i) \quad (9)$$

$$M = \sqrt{MV_x^2 + MV_y^2} \quad (10)$$

where $M(n,i)$ denotes the magnitude of the motion vector in a PU within the $i^{th}$ CU of the $n^{th}$ frame and where $C$ corresponds to the total number of CUs in the $n^{th}$ frame. Subscripts $x$ and $y$ correspond to the coordinates $(x,y)$ of the motion vector $MV$. We increment $q$ because preliminary experiments revealed that increasing $q$ by values above 1, especially for high initial QP values, can incur an increase in conspicuous block boundary compression artifacts.

Value $q$ is computed as follows:

$$q = p \times \ln(\lambda) + z \quad (11)$$

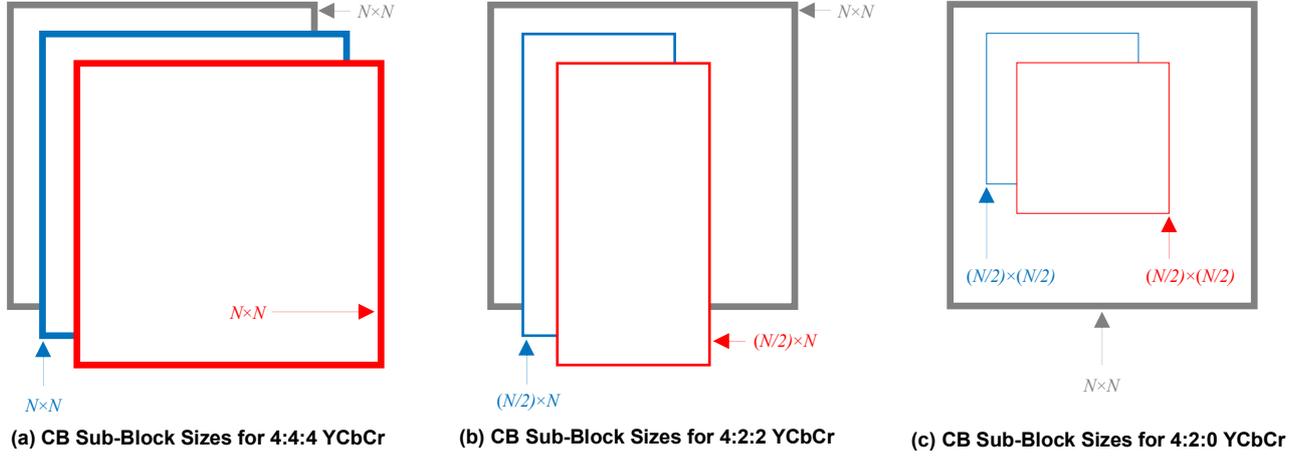

**Fig. 2.** Sizes of CB sub-blocks in a 2$N$×2$N$ CU: Y (gray), Cb (blue), Cr (red). In C-BAQ, there are four constituent sub-blocks of the Y, Cb and Cr CBs in a 2$N$×2$N$ CU. Each subfigure specifies the size of CB sub-blocks for different input video data: (a) for 4:4:4 YCbCr video data, the sub-block sizes for Y, Cb and Cr are all $N$×$N$, (b) for YCbCr 4:2:2 video data, the sub-block sizes are as follows: Y = $N$×$N$, Cb = ($N$/2)×$N$ and Cr = ($N$/2)×$N$, (c) for YCbCr 4:2:0 video data, the sub-block sizes are as follows: Y = $N$×$N$, Cb = ($N$/2)×($N$/2) and Cr = ($N$/2)×($N$/2).

where $p$ is the slope, $z$ denotes the intercept, ln corresponds to the natural logarithm and $\lambda$ is lambda. As specified in [7], $p$ = 4.2005 and $z$ = 13.7122. Lambda $\lambda$ is computed in (12):

$$\lambda = H \times W_k \times 2^{\frac{QP-12}{3}} \quad (12)$$

where $H$ is a parameter related to the coding structure, $QP$ is the frame level QP value and $W_k$ corresponds to a weighting factor that depends on the encoding configuration of the QP offset hierarchy level of the current picture within the Group Of Pictures (GOP) structure. Therefore, the values of k and, thus, $W_k$ are contingent upon the QP offset hierarchy level and the slice type. The QP hierarchy level is related to rate distortion cost functions and the Lagrangian constant values. If, for example, the All Intra encoding configuration is used, k=0, the QP offset adjustment $OA$=0 and $W_k$=0.57. When the Random Access configuration is used (B slice based configuration), k=4, $OA$=3 and $W_k$=0.68×min(2.0,4.0,($QP$–12)/6.0) [1]. $H$ is computed in (13):

$$H = \begin{cases} 1.0 - \min(0, 0.5, 0.005 \times BF) & \text{for } RP; \\ 1.0 & \text{for } NRP. \end{cases} \quad (13)$$

where $BF$ is equal to the number of B frames, $RP$ stands for referenced pictures and $NRP$ stands for non-referenced pictures.

Value $X$ is computed as follows:

$$X = \frac{(s \times (Y' + Cb' + Cr')) + t_p}{(Y' + Cb' + Cr') + (s \times t_p)} \quad (14)$$

where parameters $Cb'$ and $Cr'$ correspond to the chroma Cb and chroma Cr spatial activity of a 2$N$×2$N$ CU. Recall that $s$ and $t_p$ are defined in equations (2) and (3). Parameters $Cb'$ and $Cr'$ are computed in (15) and (16), respectively.

$$Cb' = 1 + \min(Cb^2(sb)) \quad \text{where } sb = 1,...,4 \quad (15)$$

$$Cr' = 1 + \min(Cr^2(sb)) \quad \text{where } sb = 1,...,4 \quad (16)$$

Let us recall equations (4), (5) and (6), which correspond to the luma sample variance for an $N$×$N$ CU sub-block. $Cb^2(sb)$ and $Cr^2(sb)$ are quantified in exactly the same manner as $Y^2(sb)$ (i.e., the variance of chroma pixel intensities). Consequently, $Cb^2$ and $Cr^2$ are computed in (17) and (18), respectively:

$$Cb^2 = \frac{1}{N^2} \sum_{j=1}^{N^2} (BS_j - u)^2 \quad (17)$$

$$Cr^2 = \frac{1}{N^2} \sum_{j=1}^{N^2} (RS_j - v)^2 \quad (18)$$

where $BS$ in (17) and $RS$ in (18) correspond to the chroma Cb and chroma Cr sample values, respectively. Variables $u$ and $v$ denote that mean Cb and Cr sample values, as computed in (19) and (20), respectively.

$$u = \frac{1}{N^2} \sum_{j=1}^{N^2} BS_j \quad (19)$$

$$v = \frac{1}{N^2} \sum_{j=1}^{N^2} RS_j \quad (20)$$

In terms of signaling to the decoder, as with the initial QP values, the CU level QPs in both the proposed ACUQ technique and the AdaptiveQP tool are signaled in the bitstream in the Picture Parameter Set (PPS) [10].

## 4. EXPERIMENTAL EVALUATIONS & DISCUSSION

ACUQ is compared with the AdaptiveQP tool (reference anchor). We integrate ACUQ into HEVC HM 16.7 [11] and evaluate it using the QPs 22, 27, 32 and 37, in line with the Common HM Test Conditions and Software Reference Configurations [12]. Each sequence is tested using the Random Access (RA) encoding configuration using the Main 4:4:4, Main 4:2:2, Main 4:2:0 and 4:0:0 profiles. The RA encoding configuration most accurately reflects real life video deployment situations, such as broadcasting and online streaming [13].

The following JCT-VC test sequences are employed in the experimental evaluation: FourPeople (HD 720p), KristenAndSara (HD 720p), ParkScene (HD 1080p) and Traffic (UHD 1600p). Each sequence has four different versions; that is, one each for the YCbCr 4:4:4, 4:2:2, 4:2:0 and 4:0:0 formats (16 sequences in total). The bit depth for these sequences are as follows: FourPeople (8-bit for all versions), KristenAndSara (8-bit for all versions), ParkScene (8-bit for the 4:0:0 and 4:2:0 versions, and 10-bit for the 4:2:2 and 4:4:4 versions), Traffic (8-bit for the 4:0:0 and 4:2:0 versions, and 10-bit for the 4:2:2 and 4:4:4 versions). Note that the official Format Range Extension (RExt) 4:2:2 and 4:4:4 ParkScene and Traffic sequences use a minimum of 10-bits per component. We employ the FourPeople and KristenAndSara sequences because they consist of low spatial activity and low motion data. Conversely, we utilize the ParkScene and Traffic sequences because they contain high spatial activity and high motion data. We resampled the 8-bit 4:2:0 version of the FourPeople and KristenAndSara sequences to 4:2:2 and 4:4:4 for the corresponding tests. Finally, all four sequences used in the 4:0:0 simulations (the 8-bit version of each sequence) are downsampled from 4:2:0 to 4:0:0.

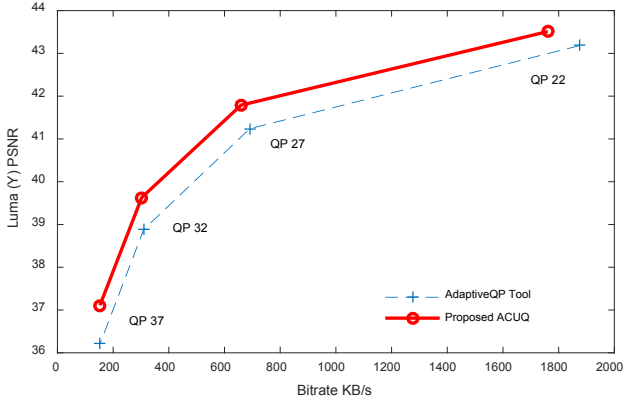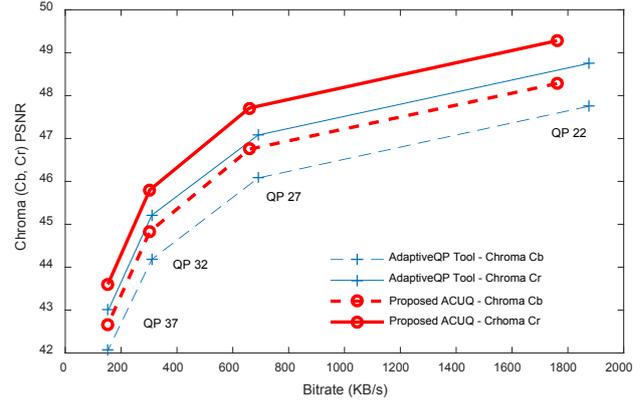

**Fig. 3.** Y PSNR improvements of the proposed ACUQ technique compared with the AdaptiveQP tool on 8-bit sequence *KristenAndSara* using the Main 4:2:2 RExt profile and the Random Access encoding configuration.

**Fig. 4.** Cb and Cr PSNR improvements of the proposed ACUQ technique compared with the AdaptiveQP tool on 8-bit sequence *KristenAndSara* using the Main 4:2:2 RExt profile and the Random Access encoding configuration.

**Table 1.** Table showing the BD-Rate improvements of the proposed ACUQ technique compared with reference anchor (AdaptiveQP tool) for the 4:0:0, 4:2:0, 4:2:2 and 4:4:4 sequences. For both the BD-Rate results and the running time results, negative percentages indicate performance improvements of the proposed ACUQ method, which are highlighted in green. The abbreviations in the Runtimes section, ET and DT, correspond to Encoding Times and Decoding Times, respectively.

| ACUQ versus Adaptive QP (YCbCr 4:0:0) | | | | | | ACUQ versus Adaptive QP (YCbCr 4:2:0) | | | | | |
|---|---|---|---|---|---|---|---|---|---|---|---|
| Sequence | BD-Rate % | | | Runtimes % | | Sequence | BD-Rate % | | | Runtimes % | |
| | Y | Cb | Cr | ET | DT | | Y | Cb | Cr | ET | DT |
| FourPeople (8-bit) | −6.9 | N/A | N/A | −2.2 | −0.6 | FourPeople (8-bit) | −13.2 | −15.6 | −16.9 | −2.0 | 0.2 |
| KristenAndSara (8-bit) | −4.2 | N/A | N/A | −0.9 | 6.1 | KristenAndSara (8-bit) | −22.4 | −27.9 | −24.6 | −0.8 | −0.4 |
| ParkScene (8-bit) | −3.0 | N/A | N/A | −1.5 | −0.4 | ParkScene (8-bit) | −6.5 | −15.2 | −16.0 | −0.9 | 3.0 |
| Traffic (8-bit) | −3.4 | N/A | N/A | −1.2 | −1.2 | Traffic (8-bit) | −4.8 | −13.4 | −17.9 | −0.8 | 0.2 |
| ACUQ versus Adaptive QP (YCbCr 4:2:2) | | | | | | ACUQ versus Adaptive QP (YCbCr 4:4:4) | | | | | |
| Sequence | BD-Rate % | | | Runtimes % | | Sequence | BD-Rate % | | | Runtimes % | |
| | Y | Cb | Cr | ET | DT | | Y | Cb | Cr | ET | DT |
| FourPeople (8-bit) | −13.5 | −18.6 | −18.5 | −0.3 | 0.0 | FourPeople (8-bit) | −14.4 | −15.8 | −16.4 | −0.8 | 0.7 |
| KristenAndSara (8-bit) | −23.1 | −26.7 | −25.2 | −0.6 | 0.5 | KristenAndSara (8-bit) | −22.0 | −24.6 | −22.9 | −1.2 | 0.7 |
| ParkScene (10-bit) | −6.4 | −15.7 | −17.2 | −1.7 | −1.1 | ParkScene (10-bit) | −7.4 | −14.3 | −17.2 | −1.0 | −1.0 |
| Traffic (10-bit) | −4.2 | −12.8 | −18.4 | −4.4 | 0.2 | Traffic (10-bit) | −4.0 | −10.7 | −16.8 | −0.7 | 2.3 |

### 4.1. BD-Rate Results

As shown in Fig. 3, Fig. 4 and Table 1, considerable coding efficiency improvements are attained by the proposed ACUQ technique in comparison with the AdaptiveQP tool. The most significant improvements, as quantified by BD-Rate reductions, are as follows: −23.1% (Y), −26.7% (Cr) and −25.2% (Cb) for the 4:2:2 KristenAndSara sequence using the Main 4:2:2 RExt profile and the Random Access configuration. Note that the proposed method consistently produces outstanding BD-Rate improvements for the 4:2:0, 4:2:2 and 4:4:4 versions of the KristenAndSara sequence. Furthermore, as confirmed in Table 1, for the 4:4:4 version of this sequence our method achieves the following coding efficiency improvements: −22.0% (Y) −24.6% (Cb) and −22.%9 (Cr) using the Main 4:4:4 RExt profile. Similarly, for the 4:2:0 version of this sequence, our method achieves the following BD-Rate reductions: −22.4% (Y) −27.9% (Cb) and −24.6% (Cr) using the Main profile (see Fig. 4).

### 4.2. Motion Component and 4:0:0 Evaluations

Although YCbCr 4:0:0 video data is not frequently utilized in contemporary video coding or televisual viewing situations, 4:0:0 video data — for example, popular monochrome movies that were created before the advent of color TV — is still viewed recreationally by the general public. In terms of the tests, because the AdaptiveQP tool in HM accounts for CU luma activity only, these 4:0:0 simulations provide us with the opportunity to assess the efficacy of the motion aspect of the proposed ACUQ technique in combination with the lambda QP refinement technique in [7]. As confirmed in Table 1, ACUQ achieves significant coding efficiency improvements for all sequences; all sequences tested have a bit depth of 8-bits. The most noteworthy improvement is attained with the FourPeople sequence; ACUQ achieves a coding efficiency improvement of −6.9%. In addition, ACUQ achieves a BD-Rate reduction of −4.9% on the KristenAndSara sequence.

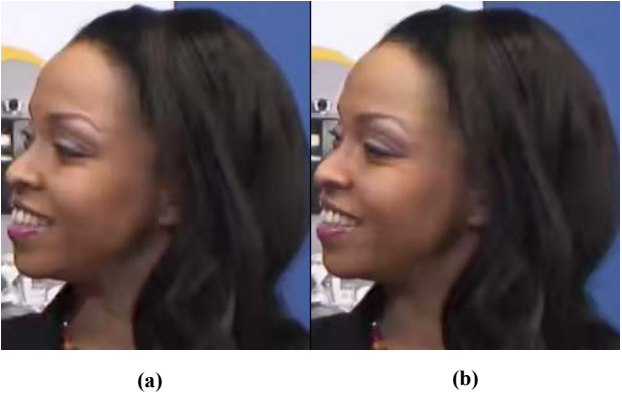

**Fig. 5.** Decoded frame from the *KristenAndSara* 4:2:0 8-bit sequence at QP 37 encoded using (a) the AdaptiveQP tool, and (b) ACUQ.

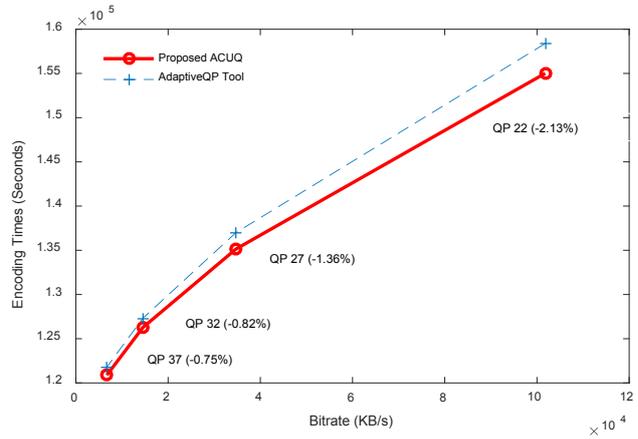

**Fig. 6.** Encoding time performance across all sequences and data points (i.e., QPs 22, 27, 32 and 37) with respect to the bitrate.

### 4.3. Runtime Performance

In the evaluation, the encoding time performance of ACUQ proved to be consistently superior in comparison with the AdaptiveQP tool (see Fig. 6). Testing ACUQ with the Traffic 4:2:2 sequence produces the most significant improvement with an encoding time reduction of −4.4% (40516.47 seconds versus 42373.41 seconds). With respect to decoding times, virtually no differences are recorded; a reduction of −1.2% is achieved for the Traffic 4:0:0 simulations (84.26 seconds versus 85.31 seconds). Overall, ACUQ attains the best decoding time performances in the 4:0:0 simulations.

### 4.4. Discussion

In the evaluation, the proposed ACUQ method outperforms the AdaptiveQP tool in terms of coding efficiency and encoding time reductions. The most notable BD-Rate reductions are achieved in the KristenAndSara and FourPeople simulations (in all tests). It is reasonable to infer that ACUQ produced a superior performance due to the technique's accounting for the variance of Y, Cb and Cr pixel intensities in CU sub-blocks, as opposed to accounting for the Y component only. Even with significant chroma subsampling (i.e., 4:2:0 — see Fig. 5), the proposed method achieves high BD-Rate reductions for both the KristenAndSara and FourPeople sequences. In essence, the synergy in terms of accounting for Y, Cb and Cr variances in CB sub-blocks, temporal masking and refining the QP according to the Lagrange multiplier [7] produces high coding efficiency gains in addition to PSNR improvements.

As regards runtimes, our integration of the lambda QP refinement technique in [7] enables the proposed ACUQ to operate effectively while negating the requirement for multiple QP optimizations in the RDO process, thus producing a notable reduction in encoding times for all tests.

### 5. CONCLUSION

A novel CU level adaptive quantization technique for HEVC, named ACUQ, is proposed. ACUQ accounts for both luma and chroma CBs in a CU — in terms of computing the variance of a CU — in addition to accounting for the temporal activity of a CU. We implemented ACUQ into HM 16.7 to evaluate and compare our method with the AdaptiveQP technique. The evaluations reveal that ACUQ achieves significant coding efficiency improvements of over 23% for the luma component and over 25% for the chroma components. Improved encoding times are also attained (a maximum 4.4% reduction).

### REFERENCES

[1] K. McCann, C. Rosewarne, B. Bross, M. Naccari, K. Sharman and G. J. Sullivan (Editors), "High Efficiency Video Coding (HEVC) Test Model 16 (HM 16) Encoder Description", *JCT-VC Document JCTVC-R1002*, Sapporo, Japan, 2014.

[2] M. Naccari and M. Mrak, "Intensity Dependent Spatial Quantization with Application in HEVC," *IEEE Int. Conf. Multimedia and Expo*, San Jose, CA, 2013, pp. 1-6.

[3] L. Prangnell, V. Sanchez and R. Vanam, "Adaptive Quantization by Soft Thresholding in HEVC," *IEEE Picture Coding Symposium*, Queensland, Australia, 2015, pp. 35-39.

[4] ITU-R, Recommendation BT.2020-2, "Parameter values for ultra-high definition television systems for production and international programme exchange," 2015.

[5] S. W. Cheadle and S. Zeki, "Masking within and across visual dimensions: Psychophysical evidence for perceptual segregation of color and motion." *Visual Neuroscience*, vol. 28, no. 5, pp. 445-51, 2011.

[6] O. Braddick, "The masking of apparent motion in random-dot patterns." *Vision Research*, vol. 13, pp. 355–369, 1973.

[7] B. Li, H. Li, J. Xu, D. Zhang, "QP refinement according to Lagrange multiplier for High Efficiency Video Coding," *IEEE Int. Symp. on Circuits and Systems*, pp. 477-480, 2013.

[8] G. Sullivan, J-R. Ohm, W. Han and T. Wiegand, "Overview of the High Efficiency Video Coding (HEVC) Standard," IEEE *Trans. Circuits Syst. Video Technol.*, vol. 22, no. 12, pp. 1649-1668, 2012.

[9] ITU-T, Recommendation H.265 (version 3) | ISO/IEC 23008-2, Information technology – Coding of audio-visual objects," *ITU-T/ISO/IEC*, 2015.

[10] V. Sze, M. Budagavi and G. J. Sullivan, "HEVC Transform and Quantization," *in High Efficiency Video Coding (HEVC): Algorithms and Architecture*, Springer International Publishing, 2014, pp. 141-170.

[11] Joint Collaborative Team on Video Coding. JCT-VC HEVC Reference Software, HM 16.7 [Online]. Available: http://hevc.hhi.fraunhofer.de/

[12] F. Bossen, "Common Test Conditions & Software Reference Configurations," *JCT-VC Document JCTVC-L1100*, Geneva, 2013, pp. 1–4.

[13] M. Wein, "Temporal Coding Structures," in *High Efficiency Video Coding - Coding Tools & Specification*, Springer International Publishing, 2015, pp. 102-114.